\documentclass{ws-procs975x65}            

\usepackage{xcolor}
\usepackage[colorlinks=true, pdfstartview=FitV,linkcolor=blue, citecolor=blue, urlcolor=blue]{hyperref}

\usepackage{psfrag}
\usepackage{latexsym,array,theorem,mathrsfs,subfigure,bm,float}
\usepackage{amsmath}
\usepackage{amsfonts}
\usepackage{amssymb}

\newcommand{\bs}{\boldsymbol}

\newcommand{\bea}{\begin{eqnarray}}
\newcommand{\eea}{\end{eqnarray}}
\newcommand{\be}{\begin{equation}}
\newcommand{\ee}{\end{equation}}

\newcommand{\mmu}{\mu}

\begin{document}
\title{Hairy black holes in the XX-th and XXI-st centuries}

\author{Mikhail S. Volkov}

\address{Laboratoire de Math\'{e}matiques et Physique Th\'{e}orique CNRS-UMR 7350,\\
Universit\'{e} de Tours, Parc de Grandmont, 37200 Tours, France.\\
Department of General Relativity and Gravitation, Institute of Physics,\\
Kazan Federal University, Kremlevskaya street 18, 420008 Kazan, Russia.\\
$^*$E-mail: volkov@lmpt.univ-tours.fr\\
}

\begin{abstract}
This is a  brief summary of the most important  hairy black hole solutions in 3+1 spacetime 
dimensions discovered over the last 25 years.
These were first of all the Einstein-Yang-Mills black holes and their various generalizations including the Higgs field,
the dilaton and the curvature corrections, and also the Skyrme black holes. More recently, 
these were black holes  supporting a scalar field violating the energy conditions or non-minimally
coupled to gravity, and also spinning black holes with 
massive complex scalar hair. 
Finally, these were black holes with  massive graviton hair.

\end{abstract}

\keywords{black holes, no-hair conjecture}

\bodymatter

\section{Introduction}
The famous no-hair conjecture formulated almost half a century ago is still a hot research topic. 
It essentially states that all black holes in Nature are of the Kerr-Newman type, and 
for a long time  this statement was broadly considered to be true. 
Besides,  it was supported by a number of mathematical results -- the 
uniqueness theorems in the Einstein-Maxwell theory and the  no-hair theorems 
proven for a number of other field-theory models. 

However, 20 years later the first manifest counter-example to the no-hair conjecture was found 
in the context of the gravity-coupled Yang-Mills theory. This discovery 
triggered an avalanche of similar findings during the last decade of the past XX-th century,
when it became clear that hairy black holes generically exist in systems with non-Abelian fields. 
As a result, one can say that, strictly speaking, the no-hair conjecture is  incorrect. 
At the same time, in most cases hairy black holes with non-Abelian fields are either 
microscopically small or unstable and when become large or  perturbed
they loose their hair. Therefore, the conjecture essentially applies to large black holes
that are astrophysically relevant. 

A new interest towards hairy black holes has emerged in the current XXI-st century,
but the focus is now shifted from gravitating non-Abelian fields
towards gravitating scalars. This is a consequence of the discovery of the dark energy,
which  can probably be modelled by a scalar field. 
If this field violates the energy conditions 
or couples non-minimally to gravity, 
then the standard  no-hair theorems do not apply and
there could be hairy black holes. 
Therefore, one is currently interested in black holes with exotic scalar fields,
such as phantoms, Galileons, etc. 
Surprisingly, this has lead to an astonishing discovery also within the conventional 
model of a complex  massive scalar field that was shown to admit spinning hairy black holes. 
Yet one more theory that has recently become popular, also in connection with the 
dark energy problem, is the ghost-free massive gravity, 
and there are hairy black holes in this theory too. 

In view of all this, this text presents an attempt to briefly summarize 
the most important asymptotically flat 
hairy black hole solutions in 3+1 spacetime dimensions 
found after their first discovery 
more than 25 years ago. This subject is  so vast that it is hardly possible to
be complete or merely objective, but at least some idea of it should be given 
by what follows.

\section{No-hair conjecture}
In 1969 Ruffini and Wheeler summarized the progress 
in black hole physics of the time in the famous phrase: 
{\it  black holes have no hair}  \cite{Ruffini:1971bza} . This means that   
\begin{itemize} 
\item 
\textcolor{black}{
All stationary black holes are completely characterized by their mass,
angular momentum, and electric charge seen from far away  in the form
of Gaussian fluxes.}
\item
Black holes cannot support \textcolor{black}{hair} = any other 
{\it independent} parameters not seen from far away. 
\end{itemize}
Therefore, 
according to this ho-hair conjecture, 
the only allowed characteristics of stationary black holes are 
those associated with the Gauss law. 
The logic behind this is the following.
Black holes are formed in the gravitational collapse, which 
is so violent a process  that it breaks all 
conservation laws not related to the exact symmetries.   
For example, the chemical content, atomic structure, baryon number, etc. are not conserved
during the collapse -- the black hole `swallows' and looses all the memory of them.  
Only few exact local symmetries, 
such as the local Lorentz or local $U(1)$, can survive
the gravitational collapse. Associated to them conserved quantities -- the mass $M$,
angular momentum $J$, and electric charge $Q$ -- cannot be absorbed by the  
black hole but remain attached to it as parameters\footnote{Black holes can also carry a magnetic charge,
if it exists.}.
They give rise 
to the Gaussian  fluxes that can be measured at infinity. 
This implies that black holes are very simple objects -- they are characterized by only three parameters,
and black holes with the same $M,J,Q$ are identically equal.

The only known in 1969 black holes  were the Kerr-Newman 
solutions. They are characterized by precisely  three parameters $M,J,Q$, 
in perfect agreement with the {conjecture}.
Moreover, a chain of {\it uniqueness theorems} 
initiated by Israel  \cite{Israel:1967wq,Robinson:1975bv,Mazur:1982db} (see \cite{heusler1996black}
for a review) 
established that the Kerr-Newman solutions describe all 
possible  stationary  {\it electrovacuum}  black holes with a non-degenerate horizon\footnote{Stationary solutions 
with zero surface gravity comprise the Israel-Wilson family of multi-black holes \cite{Israel:1972vx}.}. 
This proves the no-hair conjecture {within the Einstein-Maxwell theory}. 

Of course, not everything can be described by the Einstein-Maxwell theory, 
and the conjecture does not actually require all black holes in Nature to be necessarily Kerr-Newman. 
However, it requires them, whatever they are,
to be  completely  characterized by their charges. 
Only the {\it non-uniqueness} -- existence of  different black holes with exactly the same  
Gaussian charges   would contradict the conjecture. 
Therefore, to see if the conjecture applies to other field theories, 
one should  study the corresponding black hole solutions.

Consider  a gravity-coupled field or a system of fields of any spin 
collectively denoted by $\Psi$.  
The corresponding field equations 
together with the Einstein equations read schematically 
\be                             \label{1a}
G_{\mu\nu}=\kappa \,T_{\mu\nu}(\Psi),~~~~~~~\Box\Psi=U(\Psi).
\ee
One can wonder if these equations admit black hole solutions. 
According to the conjecture, there should be either no such solutions at all,
or only solutions labelled by their charges. 
In view of this, a number of the {\sl no-hair theorems}, largely  due to  Bekenstein
\cite{
Chase,%
Bekenstein:1971hc,%
Bekenstein:1972ky,%
Bekenstein:1972ny,%
Teitelboim:1972qx%
},
have been proven to confirm  the absence of hairy black hole solutions 
of Eqs.(\ref{1a}) in  cases where  $\Psi$ denotes  
a free scalar, spinor, massive vector, etc. field. 
The common feature in all these cases is that if $\Psi$ does not vanish, then 
the field equations require that it should 
diverge at the black hole horizon, where the curvature would diverge too.  
Notably, this happens for the massive fields. 
Therefore, to get regular black holes one is bound to set $\Psi=0$, 
but then the solution is a vacuum black hole belonging to the Kerr-Newman family. 
Similar no-go results  can be established also for the interacting fields 
\cite{
Bekenstein:1995un,%
Mayo:1996mv,%
Bekenstein:1996pn,%
Heusler:1996ft%
}. All this was confirming  the non-existence of hairy black holes. 

A peculiar finding was made in 1972 in the context of the theory with a free {\it conformally-coupled} scalar field, 
\be                        \label{B}
{\cal L}=\frac{1}{4}\,R-\frac12\,(\partial\Phi)^2-\frac{1}{12}\,R\,\Phi^2\,.
\ee
This theory admits a static (BBMB) black hole solution 
\cite{BBM,Bekenstein:1974sf},
\be
ds^2=-\left(1-\frac{M}{r}\right)^2 dt^2+\frac{dr^2}{(1-M/r)^2}+r^2d\Omega^2,~~~~~\Phi=\frac{\sqrt{3}M}{r-M},
\ee
where $d\Omega^2=d\vartheta^2+\sin^2\vartheta\, d\varphi^2$. 
The metric is extreme Reissner-Nordstrom but the scalar field 
is non-trivial, which looks like  hair. 
However, this is merely {\it secondary hair} without 
own independent parameters,  the only free parameter of the solution being
the black hole mass $M$.  Hence, the uniqueness is preserved and the no-hair conjecture holds.

The first explicit example of black holes 
whose geometry is not Kerr-Newman 
was found by Gibbons in 1982 \cite{Gibbons:1982ih} (see also \cite{Garfinkle:1990qj}) 
in the context of a supergravity model 
whose simplest version is the Einstein-Maxwell-dilaton theory, 
\be                                   \label{dil}
{\cal L}=\frac{1}{4}\,R-\frac12\,(\partial\Phi)^2-\frac14\,e^{2\Phi}F_{\mu\nu}F^{\mu\nu}\,.
\ee
The solution is obtained by setting  $F=P\sin\vartheta \,d\vartheta\wedge d\varphi$ and
\be
ds^2=-Ndt^2+\frac{dr^2}{N}+e^{2\Phi}r^2d\Omega^2,~~~~~~N=1-\frac{2M}{r},~~~~~~
e^{2\Phi}=1-\frac{P^2}{Mr}\,,
\ee
it describes static black holes with a purely magnetic Maxwell field and a non-trivial 
scalar field. The independent parameters are the 
black hole mass $M$ and  the  magnetic charge $P$, 
both subject to the Gauss law, 
while 
the scalar field carries no extra  parameters, hence it is again secondary hair%
\footnote{One can generalize the solution to include a third parameter, the asymptotic value
$\Phi(\infty)$, but this parameter would relate not to the black hole itself and rather to the 
surrounding 
world.}.

The first indication of  independent {\it primary} hair on black holes was found by Luckock and Moss in 1986 
in the context of the gravity-coupled Skyrme model \cite{Luckock:1986tr}.  
This can be viewed as a theory of three real scalars 
$\Phi^a$ defining a unitary matrix $U=\exp\{i\tau_a\Phi^a\}$ (here $\tau^a$ are the Pauli matrices)  
with the Lagrangian
\be                                             \label{Skyrme}
{\cal L}=\frac{1}{2\kappa}\,R-\frac{\alpha}{2}\,{\rm tr}\left(\partial_\mu U\partial^\mu U^\dagger\right)
-\frac{\beta}{4}\,{\rm tr}
\left(
{\cal F}_{\mu\nu}{\cal F}^{\mu\nu}
\right)
\ee
where ${\cal F}_{\mu\nu}=\left[U^\dagger\partial_{\mu}U,U^\dagger  \partial_{\nu}U\right]$. 
If the gravitational coupling $\kappa$ is small then the backreaction of the scalars on the spacetime geometry 
should also be small, hence the scalars can be obtained by solving the scalar field equations on a fixed 
Schwarzschild background.  Assuming a static and spherically symmetric ansatz
 $\Phi^a=f(r)\,n^a$ where $n^a$ is the normal to the sphere, 
 Luckock and Moss  found 
 a regular solution for $f(r)$ on the Schwarzschild background. 
This can be viewed as a perturbative 
approximation for a hairy black hole solution in the limit where the backreaction is negligible. 
A numerical evidence for a fully back-reacting solution was  reported by  Luckock \cite{Luckock:1986em}, 
 but this work was published only in the conference proceedings  and 
has gone unnoticed. 

\section{XX-th century hairy black holes}
The first broadly recognized example of a manifest violation of the no-hair conjecture 
was found  by Volkov and Gal'tsov in 1989 \cite{Volkov:1989fi,Volkov:1990}.
This result, soon confirmed by other groups   \cite{Kunzle,Bizon:1990sr}, 
was obtained within the Einstein-Yang-Mills (EYM) theory with gauge group $SU(2)$
defined by the Lagrangian 
\bea                    \label{EYM}
 {\cal L}&=&\frac{1}{4}\,R-\frac{1}{4}\,F^a_{\mu\nu}F^{a\mu\nu} ~~~~~~~~\mbox{with}~~~~~~
F^a_{\mu\nu}=\partial_\mu A_\nu^a-\partial_\nu A_\mu^a+\epsilon_{abc} A^b_\mu A^c_\nu\,.
\eea
This theory was natural to consider, since the no-hair conjecture allows  black holes
to have charges associated with local internal symmetries, but the latter can be non-Abelian.
As first observed by Yasskin \cite{Yasskin:1975ag}, 
any electrovacuum black hole can be embedded into 
the EYM theory via multiplying its $U(1)$ gauge potential by a constant hermitian matrix. This
gives a solution of the EYM equations but the
geometry is still Kerr-Newman. One can show that the theory admits no 
other black holes with non-vanishing Yang-Mills charges, 
at least in the static and spherically 
symmetric case \cite{Galtsov:1989ip}.
However, there is still a possibility to have black holes without a Yang-Mills charge,
and it is in this way that the new solutions were found.

Making the static, spherically symmetric, and purely magnetic ansatz for the fields, 
\bea                                           \label{anz}
ds^2=-\sigma^2(r)N(r)dt^2+\frac{dr^2}{N(r)}+r^2d\Omega^2,~~~~~~
A^a_i=\epsilon_{aik}\frac{x^k}{r^2}(1-w(r)), \nonumber 
\eea
the field equations reduce to a system of three coupled ODEs for three unknown functions 
$\sigma(r),N(r),w(r)$.  To solve these equations numerically,
one assumes a regular event horizon 
at a point $r_h>0$ where $N(r_h)=0$ while $\sigma(r_h)\neq 0$. 
It turns out that for  a discrete sequence of horizon values 
$w(r_h)=w_n$ ($n=1,2,\ldots$) of the Yang-Mills amplitude 
the solution extends from $r=r_h$ towards  large $r$ and approaches
the Schwarzschild metric as $r\to\infty$.
The Yang-Mills amplitude $w(r)$ oscillates $n$ times in the  $r>r_h$ region (see Fig.\ref{fig1}), 
but in the far field zone one has
$w(r)\to(-1)^n$ and 
${F^a_{ik}\sim 1/r^3}$, hence the Yang-Mills charge is zero. 
Therefore, the Yang-Mills field is supported by the black hole 
but is not visible from far away. The solutions can be labeled by two parameters: their 
ADM mass $M$ and the integer number $n$ of oscillations of the Yang-Mills field. 
Only $M$ is measurable at infinity, but for a given $M$ there are 
infinitely many  black holes with different $n$'s whose structure 
in the near field zone is different. Therefore, 
the Yang-Mills hair is {\it primary}
as it carries an independent parameter not visible from far away,
hence the uniqueness is violated.

 \def\figsubcap#1{\par\noindent\centering\footnotesize(#1)}
\begin{figure}[h]%
\begin{center}
  \parbox{2.4in}{\includegraphics[width=2.4in]{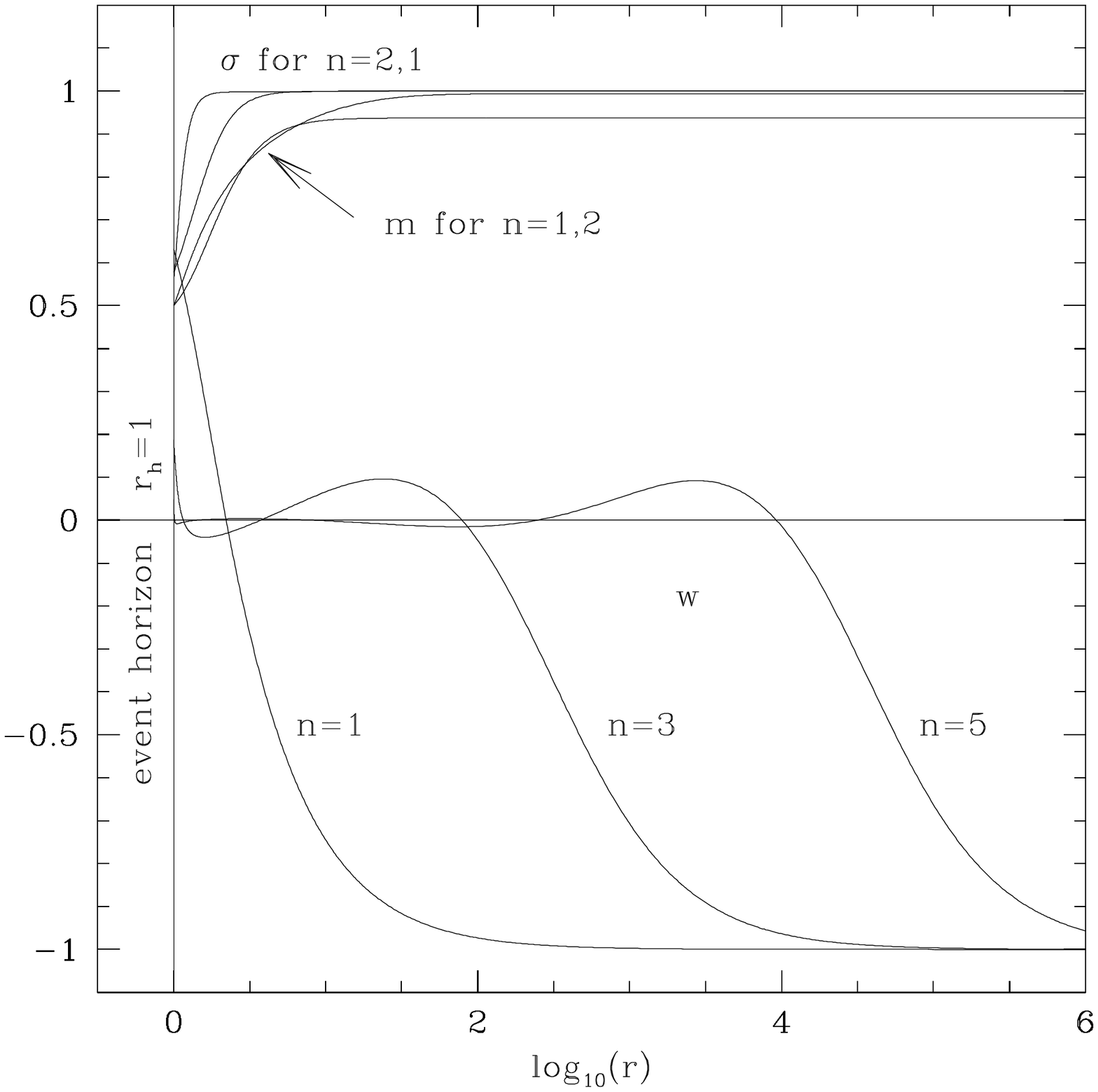}\figsubcap{a}}
  \hspace*{4pt}
  \parbox{2.4in}{\includegraphics[width=2.4in]{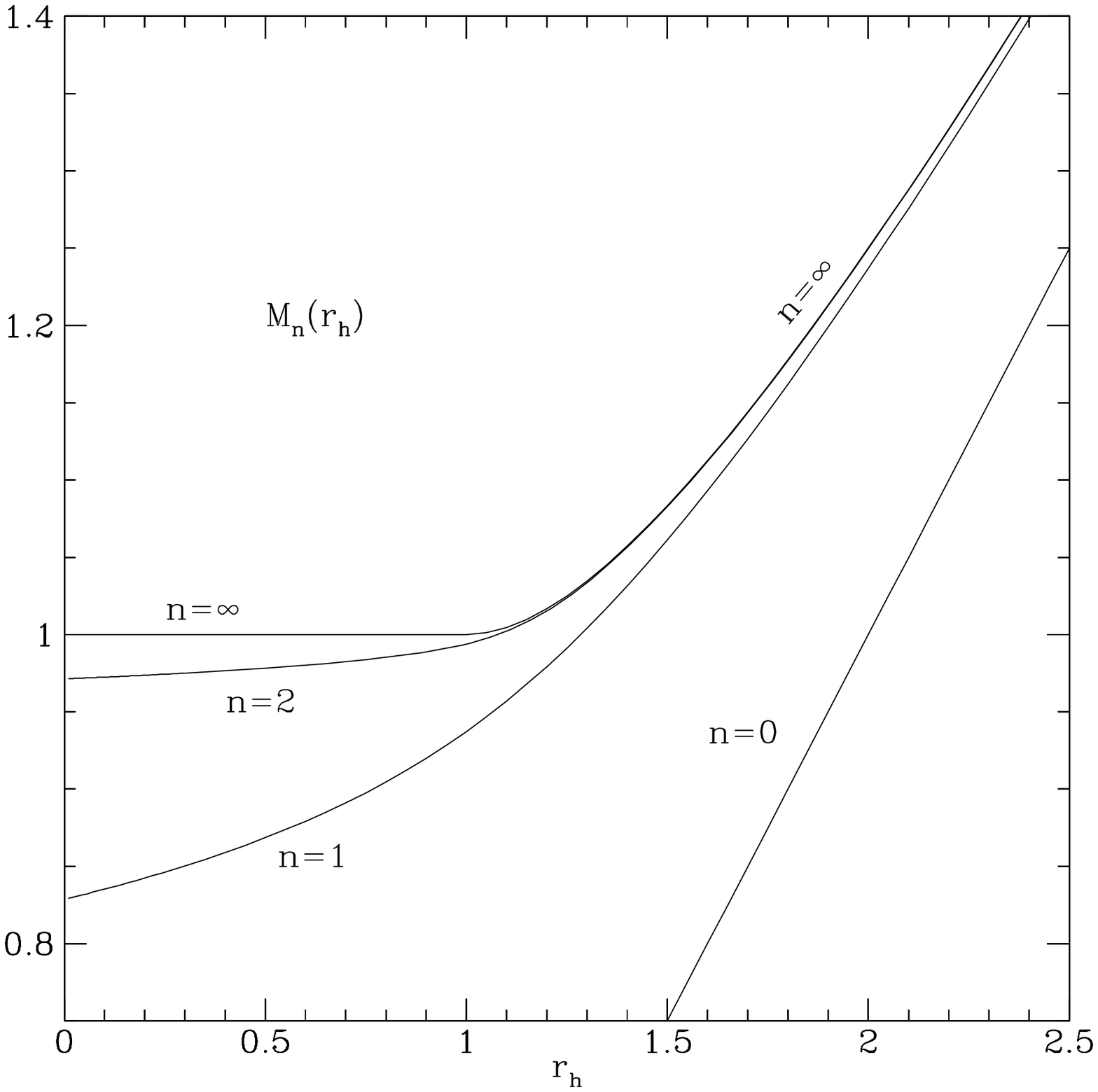}\figsubcap{b}}
  \caption{Left: profiles $w(r)$, $\sigma(r)$, $m(r)$ of the EYM black hole solutions with $r_h=1$ 
  (in Planck units);
  one has $N(r)=1-2m(r)/r$. Right: the ADM mass $M=m(\infty)$ against $r_h$.}%
  \label{fig1}
\end{center}
\end{figure}

Profiles of these EYM black holes are shown in Fig.\ref{fig1}. Their horizon size $r_h$ can be arbitrary,
and it is interesting to see that for $r_h\to 0$ the black hole masses approach  finite values corresponding 
to masses of the {\it lumps} -- globally regular particle-like objects made of gravitating Yang-Mills fields 
\cite{Bartnik:1988am}. This allows one to view the hairy black holes as non-linear superpositions 
of vacuum black holes and the lumps -- ``horizons inside classical lumps" \cite{Kastor:1992qy}.

The discovery of the EYM hairy black holes triggered an avalanche of 
similar findings in other models. The procedure 
is  similar to that in the EYM theory: one uses the metric (\ref{anz}) parameterized by
$N(r),\sigma(r)$, in addition there are matter field amplitudes. For example, in the Skyrme theory 
(\ref{Skyrme}) this is the function $f(r)$ in $\Phi^a=n^af(r)$. One has at the horizon $N(r_h)=0$ 
and one adjusts the horizon values $f(r_h)$, $\sigma(r_h)$ in such a way that the numerical solution 
matches the asymptotic values $N(\infty)=\sigma(\infty)=1$, $f(\infty)=0$. This yields black holes 
with the Skyrme hair
\cite{Droz:1991cx, Bizon:1992gb, Torii:1993vm}, or with some other hair type 
depending on the model considered. 

In most cases hairy black holes were found within various extensions of the EYM theory (\ref{EYM}). 
For example, the theory can be generalized to gauge group $SU(N)$, 
in which case black holes carry a non-zero (chromo)electric charge \cite{Baltsov:1991au},
but there is still a part of the gauge potential that  oscillates 
in the near-horizon region and is not characterized by any charge, thus violating the uniqueness.  

One can extend the EYM theory by adding a Higgs field triplet $\Phi^a$ in the adjoint representation, 
\bea                    \label{EYMH}
 {\cal L}&=&\frac{1}{2\kappa}\,R-\frac{1}{4}\,F^a_{\mu\nu}F^{a\mu\nu}-\frac12 D_\mu\Phi^a D^\mu\Phi^a
 -\frac{\lambda}{4}(\Phi^a\Phi^a-v^2)^2\,,
 \eea
 where $D_\mu\Phi^a=\partial_\mu\Phi^a+g\,\epsilon_{abc}A^b_\mu\Phi^c$ and $g,\lambda,v$
 are parameters. In flat space limit, $\kappa\to 0$, the theory admits regular soliton solutions -- magnetic 
 monopoles. For $\kappa\neq 0$ there are regular gravitating monopoles, but there are also black holes
with non-trivial Yang-Mills-Higgs hair violating the uniqueness 
\cite{Lee:1991vy,%
Breitenlohner:1991aa,%
Aichelburg:1992st%
}. Hairy black holes exist also within a global version of this theory \cite{Radu:2011uj}.
Alternatively,
the EYM theory can be extended by adding a doublet Higgs field in the fundamental 
representation, in which case one also finds gravitating solitons and hairy black holes \cite{Greene:1992fw}.

Yet another possibility to extend the pure EYM model (\ref{EYM}) is to add  terms inspired 
by string theory, for example 
\bea                    \label{EYMD}
 {\cal L}&=&\frac{1}{2\kappa}\,R-\frac12 (\partial\Phi)^2
 +e^{\gamma\Phi}\,\left(-\frac{1}{4}\,F^a_{\mu\nu}F^{a\mu\nu}+\alpha\,{\cal G}_{\rm GB}\right)-V(\Phi),
 \eea
 where $\gamma,\alpha$ are parameters and 
${\cal G}_{\rm GB}=R_{\mu\nu\alpha\beta}R^{\mu\nu\alpha\beta}
-4R_{\mu\nu}R^{\mu\nu}+R^2$ is the Gauss-Bonnet invariant. 
Setting here $\alpha=V(\Phi)=0$ gives the EYM-dilaton theory, admitting  the hairy black holes 
and regular lumps very similar to those in the pure EYM theory \cite{Lavrelashvili:1992ia}.
Setting $\alpha=0$, $\gamma=2$, $V(\Phi)=-(1/8)\exp(-2\Phi)$ gives a sector of the $N=4$  
gauged supergravity whose hairy black hole solutions  can be used for a holographic
description of confinement \cite{Gubser:2001eg}. 
One can consistently set the gauge field to zero,
 and what remains corresponds (if $V(\Phi)=0$) to the dilaton-Gauss-Bonnet model,
\bea                    \label{EYMD1}
 {\cal L}&=&\frac{1}{2\kappa}\,R-\frac12 (\partial\Phi)^2
 +\alpha\,e^{\gamma\Phi}\,{\cal G}_{\rm GB}. 
 \eea
This theory also admits hairy black holes, but the scalar hair is secondary and completely 
determined by the black hole mass 
\cite{Mignemi:1992nt,Kanti:1995vq,Torii:1996yi,Alexeev:1996vs}. 

Quite a lot has been done to study also more general hairy black holes with lower symmetry. 
This has revealed surprising facts. 
For example, although it is natural to believe the static black holes to be always spherically symmetric 
(if their horizon is non-degenerate), one finds that static EYM(-dilaton) black holes can be only 
axially  symmetric \cite{Kleihaus:1997ic} (a similar result holds for the Skyrme black holes \cite{Sawado:2003at}). 
The EYM-dilaton theory admits also stationary spinning black holes, which is not very surprising, 
however, there are  solutions which remain stationary and non-static even when their angular momentum
vanishes \cite{Kleihaus:2000kg}. Stationary spinning black holes have also been constructed 
in the dilaton-Gauss-Bonnet theory \cite{Kleihaus:2011tg}, in the Einstein-Skyrme theory \cite{Ioannidou:2006nn},
and  in the conformally-coupled model \eqref{B} \cite{Astorino:2014mda}. 

Here is a brief summary of some  common features of the XX-th century hairy black holes. 
They exist in generic models with gravity-coupled non-Abelian gauge fields and support primary 
Yang-Mills hair characterized by radial oscillations.  
They can support in addition a scalar Higgs  and/or  dilaton field. 
Static solutions may or may not be  spherically symmetric, 
while stationary solutions may be spinning but not necessarily. 
When the black hole horizon shrinks to zero, the solutions reduce not to the vacuum 
but to globally-regular ``lumps'' made of the gravitating Yang-Mills(+other) fields. 
Therefore, the hairy black holes can be viewed as horizons inside lumps. 

In models containing apart from Planck's mass also another mass scale, as for example in the EYM-Higgs theory,
the black holes  have a {finite maximal size} above which they loose their hair and reduce to 
electrovacuum black holes. 
In models with only one (Planck) mass scale, as in the EYM(-dilaton)
theory, hairy black holes can be of any size. On the other hand, solutions in models with only one mass
scale are generically unstable \cite{Straumann:1990as,Volkov:1995np}, while those  in 
the two-scale models  can be stable -- at least they do not show the most dangerous S-mode 
instability \cite{Aichelburg:1992st,Hollmann:1994fm}. The unstable solutions in the pure EYM theory can be stabilized
by adding a negative cosmological constant, which introduces a second scale
\cite{Winstanley:1998sn}. A more detailed account 
of the XX-th century black holes can be found in the review article \cite{Volkov:1998cc}.

\section{XXI-st century black holes -- solutions with a scalar field}

One discovered at the very end of the last century 
 that our universe is actually accelerating \cite{1538-3881-116-3-1009,0004-637X-517-2-565}. 
 Explaining this fact within the 
GR context requires introducing a dark energy of an unclear origin. 
Assuming it to be dynamical and not just a cosmological constant, the most popular dark energy 
models consider  a cosmic scalar field. This field 
couples to gravity, but not necessarily minimally, and its energy may  be not 
necessarily positive. This explains the current interest towards models with gravitating scalar fields,
but if one believes that they describe cosmology, then they should describe black holes as well. 

\subsection{Minimal models violating the  energy conditions}
The simplest possibility is to consider
\be                            \label{q}
 {\cal L}=\frac{1}{4}\,R -\frac{1}{2}\,(\partial\Phi)^2-V(\Phi).
 \ee
However, 
if  $V(\Phi)\geq 0$ (the strong energy condition)
 then one can show  \cite{Heusler:1996ft} that the only static and spherically 
symmetric black hole in the model is the vacuum one, with $\Phi=\Phi_0$ and $V(\Phi_0)=0$. 
Therefore,  to obtain black holes in this case
 one should abandon the strong energy condition. 
At the same time, according to the modern paradigm, the potential $V(\Phi)$ may  be not necessarily positive,
and if it is non-positive definite, 
then one finds indeed 
 asymptotically flat hairy black holes. 
 Such solutions can be constructed  numerically for a chosen 
function $V(\Phi)$ \cite{Nucamendi:1995ex}, or the potential can be specially adjusted 
to obtain solutions analytically. 
For example, choosing \cite{Bronnikov:2001ah,Zloshchastiev:2004ny,Gonzalez:2013aca,Cadoni:2015gfa} 
\be                               \label{V}
V(\Phi)=3\,\sinh(2\Phi)-2{\Phi}\left[\cosh(2\Phi)+2\,\right]
\ee
yields, as can be directly checked, an exact solution of the theory, 
\bea                          \label{Bron}
ds^2&=&-Ndt^2+\frac{dr^2}{N}+R^2\,d\Omega^2\,,~~~~~~~R^2=r(r+2Q),~~~\nonumber \\
N&=&1-4\,[
Q(Q+r)-R^2\,\Phi
],~~~
~~~e^{2\Phi}=1+\frac{2Q}{r}\,.
\eea
If $4Q^2>1$ then this describes hairy black holes since
$N(r_h)=0$ at $r_h>0$ while for $r\to\infty$ one has $N=1-8 Q^3/(3r)+\ldots$  and 
$\Phi=Q/r+\ldots$. The theory also admits the  vacuum Schwarzschild solution with $\Phi=0$, $R=r$   
and $N=1-2M/r$.  These solutions are completely characterized by their mass $M$ and the scalar charge $Q$,
since  to different $M,Q$ pairs there correspond different solutions. 
 However, only $M$ gives rise to the Gaussian flux,  
and there can be different solutions with the same $M$.  Hence, 
 the no-hair conjecture is violated.
 
The physical role of these solutions is not very clear, but  at least they are simple enough. 
Other explicitly known solutions \cite{Anabalon:2012ih,Anabalon:2013qua}  obtained via adjusting the potential $V(\Phi)$
are extremely complicated, although some of them 
may perhaps be related to a supergravity model \cite{Anabalon:2013eaa}. 
Unfortunately, all solutions of this type seem to be generically unstable under small fluctuations
\cite{,Nucamendi:1995ex,Anabalon:2013baa}.

One may violate also the weak energy condition by  considering the 
phantom model with a negative kinetic energy, 
\be                            \label{f}
 {\cal L}=\frac{1}{4}\,R +\frac{1}{2}\,(\partial\Phi)^2-V(\Phi).
 \ee
It turns out \cite{Graham:2014mda} that in this case too one needs a non-positive definite potential 
to get black holes, but these solutions \cite{Bronnikov:2005gm}
are unstable as well \cite{Bronnikov:2012ch}.

\subsection{Spinning black holes with scalar hair}

The above solutions are rather exotic, but
the analysis of the scalar models has lead to an important discovery within 
the simplest model with a massive {\it complex} scalar, 
\be                            \label{fff}
 {\cal L}=\frac{1}{4}\,R -|\partial\Phi|^2-\mu^2|\Phi|^2\,.
 \ee
 There is a no-hair theorem \cite{Pena:1997cy}
that forbids {\it static} and spherically symmetric black holes 
 in this case.  Nevertheless, Herdeiro and Radu \cite{Herdeiro:2014goa} were able to show 
 that there are {\it stationary} spinning black holes with primary scalar hair 
 which do not have the static limit. 
 Their existence can be revealed by the following means. Setting $\Phi=0$, the theory 
 admits the vacuum Kerr solution, 
 \be                          \label{K}
 ds^2=-Ndt^2+\frac{1}{\Delta}(d\varphi+Wdt)^2+R^2(dr^2+r^2d\vartheta^2),
 \ee
 where $N,\Delta,W,R$ depend on $r,\vartheta$. The solution is characterized by its mass $M$
 and angular momentum $J$. 
 Suppose $\Phi$ is so small
 that one can neglect its backreaction on the geometry, then it should fulfil the Klein-Gordon equation 
 on the Kerr background. 
 It turns out \cite{Hod:2012px} that this equation admits stationary 
 bound  state solutions  (scalar clouds) described by 
 \be                           \label{Kf}
 \Phi=F(r,\vartheta)\,\exp\{i\omega t+im\varphi\}
 \ee
 with $m=\pm 1,\pm 2,\ldots$ and  $\omega=m\Omega_H$ where $\Omega_H$ is the event horizon 
 angular velocity.  For these solutions one can compute the global Noether charge $Q$ distributed outside
 the black hole, and since $\Phi$ is assumed to be small, $Q$ is small too. These scalar clouds 
 on the Kerr background can be viewed as an approximation to the  fully back-reacting solutions described by 
 three parameters $M,J,Q$ in the limit where $Q$ is  small. The full solutions for a non-small $Q$ 
 were constructed  by Herdeiro and Radu \cite{Herdeiro:2014goa}
 within the ansatz (\ref{K}),(\ref{Kf}) by starting from the scalar clouds and then 
 iteratively increasing the scalar field amplitude, adjusting at the same time $\omega$ and the 
 event horizon size  to preserve the regularity at the horizon and at infinity. 
 
 Here are some properties of these solutions. 
  For a given value of the azimuthal number $m$ they can be 
 labeled by $M,J,Q$. Only $M$ and $J$ are seen at infinity and there are 
 different solutions with the same $M,J$, hence the no-hair conjecture is violated. 
 For $Q\to 0$ one recovers  the scalar clouds. When 
 the event horizon shrinks, the solutions reduce to globally regular gravitating solitons with $J=mQ$:
 boson stars  \cite{Schunck:2003kk}. For solutions with fixed $M$ and $Q\neq 0$ 
 the angular momentum $J$ can vary only within a finite range and there 
 is a non-zero lower bound for $J$, hence the solutions do not admit 
 the static limit $J\to 0$. On the other hand, they
 can have $J>M^2$ thus violating the Kerr bound.  
 Their mass $M$  also varies within a finite range and cannot be too large if $Q\neq 0$.
 Stability of these solutions remains an open issue. 
 

\subsection{Non-minimal models}
The considered above examples assume the minimal 
coupling of the scalar to gravity. Some of them can be generalized to the non-minimally coupled 
case in the standard way,
by adding to the 
Lagrangian the $\xi R\Phi^2$ term. 
There is, however, a much more general way to 
approach the problem. All possible gravity-coupled scalar field models 
with at most second order field equations 
are contained in the 
general theory  discovered  by Horndeski in 1972 
\cite{Horndeski:1974wa}\footnote{
There is an even more general possibility allowing for higher order equations but still keeping only three propagating 
degrees of freedom \cite{Gleyzes:2014qga}, but we do not consider it here.}.
This theory is described by 
the Lagrangian 
$
{\cal L}={\cal L}_2+{\cal L}_3+{\cal L}_4+{\cal L}_5
$
with
\bea                        \label{horn}
{\cal L}_2&=&G_2(X,\Phi),~~~~~~~~~~
{\cal L}_3=G_3(X,\Phi)\,\Box\Phi\,, \nonumber \\
{\cal L}_4&=&G_4(X,\Phi)\,R+\partial_X G_4(X,\Phi)\,\delta^{\mu\nu}_{\alpha\beta}\,
\nabla_{~\mu}^\alpha\Phi\nabla_{~\nu}^\beta\Phi\,,  \nonumber \\
{\cal L}_5&=&G_5(X,\Phi)\,G_{\mu\nu}\nabla^{\mu\nu}\Phi
-\frac16\,\partial_X G_5(X,\Phi)\,\delta^{\mu\nu\rho}_{\alpha\beta\gamma}\,
\nabla_{~\mu}^\alpha\Phi\nabla_{~\nu}^\beta\Phi\nabla_{~\rho}^\gamma\Phi  \,,
\eea
where  $X= -\frac12 (\partial\Phi)^2$ and 
$G_k(X,\Phi)$ are arbitrary functions. 
This theory contains all previously studied models 
with a gravity-coupled scalar field.  
If the coefficient functions $G_k$ do not depend on $\Phi$, 
it reduces to the covariant Galileon theory 
 invariant under 
shifts $\Phi\to\Phi+\Phi_0$. 
 Due to this shift symmetry, the 
equation for $\Phi$ is the conservation condition for the corresponding Noether current, 
$\nabla_\mu J^\mu=0$.

The latter fact was used by Hui and Nicolis \cite{Hui:2012qt} 
to produce a no-hair theorem for Galileons in the 
static and spherically symmetric case. 
The argument is simple: the metric can always be chosen 
in the form (\ref{Bron}) while $\Phi=\Phi(r)$, hence the only 
non-zero component of the 
Noether current is  $J^r$. 
The scalar field equation is $(R^2 J^r)^\prime=0$ hence
$J^r=C/R^2$ where $C$ is an integration constant, however, $C$ should be set to zero
since otherwise the invariant $J^\mu J_\mu= C^2/(NR^4)$ would diverge at the horizon where 
$N$ vanishes. This shows that $J^r=0$ everywhere.  Now, assuming the Lagrangian 
to be  a polynomial in $\Phi^\prime$
of at least {\it second} order gives
\be                 \label{J}
J^r=N\frac{\partial{\cal L}}{\partial\Phi^\prime}=\Phi^\prime\, F(\Phi^\prime,N,N^\prime,R,R^\prime,R^{\prime\prime})=0.
\ee 
For asymptotically flat solutions $F$ approaches a constant value as $R\to\infty$
hence it cannot vanish, therefore one concludes that $\Phi^\prime=0$.

However, 
Sotiriou and Zhou \cite{Sotiriou:2014pfa} pointed out that 
${\cal L}$ can also contain terms linear in $\Phi^\prime$. 
 Specifically, if
$
G_5(X)=const.\times \ln(|X|) +\tilde{G}_5(X)
$
then the Galileon theory becomes
\be                                 \label{GB}
{\cal L}=R+X+\alpha\,\Phi\,{\cal G}_{\rm GB}+\ldots
\ee
where $\alpha$ is a constant and ${\cal G}_{\rm GB}
=\nabla_\mu G^\mu$
 is the Gauss-Bonnet invariant. The dots denote terms depending on 
 the remaining coefficient functions, but  they can be set to zero for simplicity and
 then the current is $J^\mu=\partial^\mu\Phi+\alpha G^\mu$. Therefore, the zero current condition 
$J^r=N\Phi^\prime+\alpha G^r=0$ {does not} imply that $\Phi^\prime=0$. 
 As a result, one can construct asymptotically flat black holes with a regular horizon
 and a non-trivial scalar field \cite{Sotiriou:2014pfa}. 
 These solutions are very similar to those \cite{Mignemi:1992nt,Kanti:1995vq,Torii:1996yi,Alexeev:1996vs} 
 previously studied
 in the dilaton-Gauss-Bonnet model (\ref{EYMD1}) (also a Horndeski theory), 
 for example there is a non-trivial lower bound for their mass. 
Their scalar charge is not 
 an independent parameter but determined by the black holes mass, hence the hair is 
 only secondary. 
 
 Sotiriou and Zhou also 
 noticed \cite{Sotiriou:2014pfa} that symmetries of 
the problem allow for a more general choice of $\Phi$ since actually only its gradient 
needs to be static and spherically symmetric. 
This idea has been implemented by Babichev and Charmousis \cite{Babichev:2013cya} 
within the context of a  particular Galileon model, 
\be                              \label{Fab5}
{\cal L}=\mmu\, R-(\sigma\, G_{\mu\nu}+\varepsilon\, g_{\mu\nu})\nabla^\mu\Phi\nabla^\nu\Phi-2\Lambda\,.
\ee
Specifically, choosing $\Phi=Q\,t+\phi(r)$ and setting $\mu=\epsilon=\Lambda=0$ gives
an exact solution, 
\be                                              \label{SS}
ds^2=-Ndt^2+\frac{dr^2}{N}+r^2d\Omega^2,~~~~~~~\Phi=Q\,t\pm Q\int\frac{\sqrt{1-N}}{N}\,dr\,,
\ee
with $N=1-2M/r$. This is called ``stealth Schwarzschild'' because, 
although the full backreaction problem is solved, the metric is pure Schwarzschild and
the scalar field effectively does not back-react, even though 
it diverges at the (past or future) event horizon and  at infinity. 
 It is interesting to see how the no-go  argument is circumvented. 
The Noether current $J^\mu$ has now two components, $J^0$ and $J^r$, 
but $J^\mu J_\mu$ is still finite at the horizon.  
The scalar field equation still has the form (\ref{J}) and is fulfilled by setting 
$F=0$ and not $\Phi^\prime=0$, which is possible because $\partial_t\Phi\neq 0$.

Notice that $\Phi$  for this solution is non-trivial
 in the far field zone, which breaks the asymptotic Lorentz invariance. 
It is possible  \cite{Sotiriou:2014pfa} that there could be other similar 
solutions if one allows  the Lagrangian to contain 
negative powers of $\Phi^\prime$.

Similar solutions with the linear time-dependence of  $\Phi$ have been obtained 
within other subclasses of the general covariant Galileon model \cite{Kobayashi:2014eva},
as well as in the $F(R)$ gravity \cite{Zhong:2015ina} (also a Horndeski theory). 
However, such solutions seem to be generically unstable 
\cite{Ogawa:2015pea}. 

The model (\ref{Fab5}) also admits an exact solution 
\cite{Rinaldi:2012vy,Minamitsuji:2013ura,Anabalon:2013oea}
 for generic values of its parameters 
$\mu,\epsilon,\sigma,\Lambda$ and with $\Phi=\Phi(r)$,
\bea                                    \label{rinal}
ds^2&=&-Ndt^2+\frac{dr^2}{H}+r^2 d\Omega^2,~~~{H}=\frac{(\eta\, r^2+1)N}{(rN)^\prime },~~~
\Phi^{\prime 2}=\frac{\eta\,(\lambda+\mu )r^2}{\sigma\,(\eta\,r^2+1)H}\,,\nonumber \\
N &=&\eta\,(\mu-\lambda)^2 r^2+3(\mu-\lambda)(3\mu+\lambda)
+3(\lambda+\mu)^2\frac{\arctan(\sqrt{\eta}\,r)}{\sqrt{\eta}\,r}-\frac{2M}{r}\,,
\eea
where $\eta=-\varepsilon/\sigma$ and $\lambda=\Lambda/\eta$.
One can adjust the parameters $\mu,\eta,\lambda$ and the integration constant $M$ such 
that $N,H$ vanish at $r=r_h$ and are positive for $r>r_h$, which corresponds to a black hole.
This solution fulfills Eq.(\ref{J}) with $F=0$, which is possible because the geometry is not asymptotically flat. 
However, there is a  problem, 
since  if one adjusts the parameter $\sigma$ such that 
$\Phi^{\prime 2}>0$ outside the horizon, then one will have $\Phi^{\prime 2}<0$  inside
hence $\Phi$ becomes complex-valued. 
This renders unclear the status of the solution. 
More information on black holes with scalar hair can be 
found in the recent review \cite{Herdeiro:2015waa}.

\section{XXI-st century  black holes -- solutions with massive gravitons} 

Recently becoming popular
theories of massive gravity  provide 
an alternative explanation for the cosmic self-acceleration. 
The basic idea is simple -- if gravitons are massive, then the Newton potential 
is replaced by the Yukawa potential, hence the gravitational attraction is weaker
at large distances and the universe expands faster. 
Models with massive gravitons 
have been known for a long time but attracted a serious attention  only recently, 
after the discovery of the special massive gravity theory \cite{deRham:2010kj} 
circumventing the long standing problem of the ghost -- un unphysical 
mode in the spectrum rendering everything unstable
(see \cite{Hinterbichler:2011tt,deRham:2014zqa}
for a review). 

The massive gravity theory is described by two metrics, $g_{\mu\nu}$
and $f_{\mu\nu}$, of which the first one is dynamical and the second one is  fixed. 
The symmetry between the two metrics is restored in the ghost-free bigravity theory 
where they are both dynamical \cite{Hassan:2011zd}. This theory describes 
two interacting gravitons, one massive and one massless. In the limit where
the massless graviton decouples, the theory reduces to the massive gravity. 
Alongside with the self-accelerating cosmologies \cite{Volkov:2013roa}, 
this theory admits hairy black holes.

The action of the  ghost-free bigravity  is 
\bea                                      \label{1}
\frac{m^2}{M^2_{\rm Pl}}\,S
=\frac{1}{2\kappa_1}\int \, R(g)\sqrt{-{ g}}\,d^4x
+\frac{1}{2\kappa_2}\int \, R(f)\sqrt{-{ f}}\,d^4x
-\int{\cal U}\sqrt{-g}\,d^4x
\,,
\eea
where  all quantities on the right are rescaled to be dimensionless, 
the length scale being the inverse graviton mass $1/m$, and $\kappa_1+\kappa_2=1$. 
The interaction between the metrics is determined by the scalar potential
\bea                          \label{2}
{\cal U}=b_0+
b_1\sum_{a}\lambda_a
+b_2\sum_{a<b}\lambda_a\lambda_b 
+b_3\sum_{a<b<c}\lambda_a\lambda_b\lambda_c
+b_4\,\lambda_0\lambda_1\lambda_2\lambda_3\,,
\eea
where  $\lambda_a$ are eigenvalues of 
$
\bs{\gamma}^\mu_{~\nu}=
\sqrt{{{g}}^{\mu\alpha}{{f}}_{\alpha\nu}}
$ 
with
 the 
square root
understood in the sense that  
$
\bs{\gamma}^\mu_{~\alpha}\bs{\gamma}^\alpha_{~\nu}
={{g}}^{\mu\alpha}{{f}}_{\alpha\nu}.
$ 
The five parameters $b_k$ in \eqref{2} can be arbitrary, but if one requires the Minkowski space
$g_{\mu\nu}=f_{\mu\nu}=\eta_{\mu\nu}$ to be a solution of the theory, then 
$b_k=b_k(c_3,c_4)$ where $c_3,c_4$ are independent.
Varying the action with respect to $g_{\mu\nu}$
and $f_{\mu\nu}$ 
gives two coupled sets of equations.

All known bigravity black holes \cite{Volkov:2012wp,Brito:2013xaa} 
can be divided into three types. 

\noindent
I. {\it Proportional solutions: } 
$f_{\mu\nu}=\lambda^2 g_{\mu\nu}$
with 
$G_{\mu\nu}(g)+\Lambda(\lambda) g_{\mu\nu}=0$ and
\be                               \label{algebre}
\Lambda(\lambda)=\kappa_1\left[P_0(\lambda)+\lambda P_1(\lambda)\right]=
\frac{\kappa_2}{\lambda}\left[P_1(\lambda)+\lambda P_2(\lambda)\right], 
\ee
where  $P_k(\lambda)\equiv b_k+2b_{k+1}\lambda+b_{k+2}\lambda^2$.
The second equality here implies that $\lambda$ is a root of the algebraic 
equation, and depending on its choice  $\Lambda(\lambda)$ can be 
positive, negative, or zero. Therefore, solutions are the ordinary 
GR black holes, for example 
Schwarzschild or Schwarzschild-(anti)-de Sitter. 
If $b_k=b_k(c_3,c_4)$ then $\lambda=1$  is always a root 
 and $\Lambda(1)=0$, hence setting $g_{\mu\nu}=f_{\mu\nu}$ 
reduces the bigravity to the vacuum GR.

\noindent
II. {\it Non-diagonal solutions:}
\bea                                  \label{off}
ds_g^2&=&-\Sigma\, dt^2+\frac{dr^2}{\Sigma}+r^2 d\Omega^2,
~~~~~~\Sigma=1-\frac{2M_g}{r}-\frac{\Lambda_g}{3}\,r^2, \nonumber \\
\frac{1}{\lambda^2}\,ds_f^2&=&-\Delta\, {dT}^2+\frac{dr^2}{\Delta}+r^2 d\Omega^2,
~~~~~~\Delta=1-\frac{2M_f}{r}-\frac{\lambda^2\Lambda_f}{3}\,r^2, 
\eea
with $P_1(\lambda)=0$, $\Lambda_g=\kappa_1 P_0(\lambda)$, and 
$\Lambda_f=\kappa_2 P_2(\lambda)$. Both metrics are 
Schwarzschild-(anti)-de Sitter but with different cosmological terms
which do not necessarily have the same sign. 
The time $T$ for the 
second metric is not the coordinate time $t$ but a function $T(t,r)$ 
subject to 
\be			\label{eT}
\frac{\Delta}{\Sigma}\,({\partial_t T})^2+
\frac{\Delta\Sigma}{\Delta-\Sigma}\,({\partial_r T})^2=1\,.
\ee
 A simple solution of this equation can be 
obtained by separating the variables, 
\be
{T}=t+\int \frac{dr}{\Sigma}-\int\frac{dr}{\Delta}\,\equiv
t+r^\ast_\Sigma-r^\ast_{\Delta},
\ee
and using the light-like coordinate
$
V=t+r^\ast_\Sigma={T}+r^\ast_{\Delta}\,
$
both metrics can be written in the Eiddington-Finkelstein form, 
\bea 
ds_g^2=-\Sigma dV^2+2dVdr+r^2d\Omega^2\,, ~~~
\frac{1}{\lambda^2}\,ds_f^2=-{\Delta} dV^2+2dVdr+r^2d\Omega^2\,.
\eea 
Other functions $T(t,r)$ subject to  \eqref{eT} give rise to physically inequivalent solutions \eqref{off}. 
Since the f-metric becomes flat for $\kappa_2\to 0$ (if $M_f=0$), 
the solutions describe in this limit black holes in the massive gravity theory with flat reference metric.
These and their stationary generalizations  \cite{Babichev:2014tfa} 
exhaust all known massive gravity black holes (more solutions exist within the 
extended versions of the theory \cite{Tolley:2015ywa}). 

\begin{figure}[th]
\hbox to \linewidth{ \hss
\resizebox{8cm}{5cm}{\includegraphics{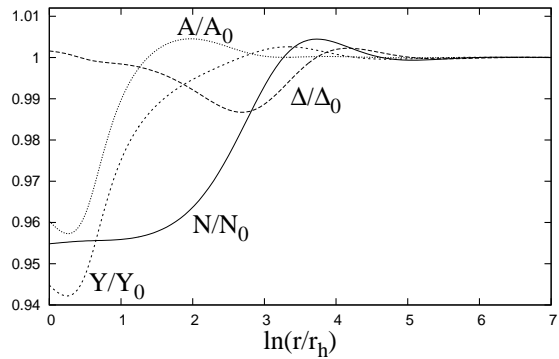}}
\hspace{1mm}
\hss}
\caption{{\protect\small 
Hairy deformations of the Schwarzschild-AdS background,  where $A_0,N_0,\Delta_0,Y_0$
relate to  the background. 
    }}
\label{fig2}
\end{figure}

\noindent
III. {\it Hairy black holes. } Black holes  of the previous two types are described by the 
known metrics. New results
are obtained when 
the two metrics are simultaneously diagonal but not necessarily proportional \cite{Volkov:2012wp}, 
\bea
ds_g^2&=&N^2dt^2-\frac{dr^2}{\Delta^2}-r^2d\Omega^2,~~~~
ds_f^2=A^2 dt^2-\frac{U^{\prime 2}}{Y^2} dr^2-U^2d\Omega^2\,, 
\eea
where 
$N,\Delta,Y ,U,A$ depend on $r$ and fulfill  a system of differential equations. 
The simplest solutions are the proportional black holes   of type I. 
More general solutions are obtained by numerically integrating the equations. 
It turns out that the equations for the three amplitudes 
$\Delta,Y,U$ comprise a closed subsystem whose local solution 
near the horizon 
\be                      \label{local}
\Delta^2=\sum_{n\geq 1}a_n(r-r_h)^n,~~
Y^2=\sum_{n\geq 1}b_n(r-r_h)^n,~~
U={ u}r_h+\sum_{n\geq 1}c_n(r-r_h)^n, \nonumber
\ee
contains {only one free parameter ${ u}$}=$U(r_h)/r_h$, that
is the ratio of the horizon size 
measured by $f_{\mu\nu}$ to that measured by $g_{\mu\nu}$. 
The horizon is common for both metrics and 
 its surface gravities and temperatures 
determined with respect to both metrics are the same. 
Choosing a value of $u$ and integrating numerically the 
equations from $r=r_h$ towards large $r$, one generically finds 
solutions approaching the proportional AdS metrics \cite{Volkov:2012wp}\footnote{
It is also possible that only one of the two metrics is asymptotically AdS \cite{Volkov:2012wp}. 
}.
These solutions describe black holes with massive graviton hair; their typical profile 
is shown in Fig.\ref{fig2}. The hair is localized in the horizon vicinity,
while far from the horizon the solutions rapidly approach the AdS. 
For a fixed horizon size $r_h$ these solutions comprise a one-parameter family labeled by $u$. 
In the $r_h\to 0$ limit
the solutions reduce to regular lumps -- localized  objects made of self-gravitating massive graviton modes. 
If on the contrary $r_h$ is very large and of the order of the inverse graviton mass, $r_h\sim 1$, then 
there are {\it discrete} values of $u$ for which the solutions are not asymptotically AdS but 
{\it asymptotically flat }\cite{Brito:2013xaa};
however this is only possible for cosmologically large black holes. 

Stability  has been studied for solutions of types I and II. The simplest 
solution of type I is the Schwarzschild black hole $g_{\mu\nu}=f_{\mu\nu}=g^{\rm (0)}_{\mu\nu}$;  
it is linearly stable in GR. 
However, when perturbed, 
$g_{\mu\nu}= g^{\rm (0)}_{\mu\nu}+\delta g_{\mu\nu},$
$f_{\mu\nu}= g^{\rm (0)}_{\mu\nu}+\delta f_{\mu\nu}$,
the perturbations $\delta g_{\mu\nu}$ and $\delta f_{\mu\nu}$ 
need not to coincide to each other, hence the GR result will not be 
necessarily recovered.  It turns out that 
the massive graviton described by the linear combination 
$h_{\mu\nu}=\delta g_{\mu\nu}+\sqrt{\kappa_2/\kappa_1  }\,\delta f_{\mu\nu}$ fulfills 
\cite{Babichev:2013una} the same equations  
as those describing the Gregory-Laflamme instability \cite{Gregory:1993vy}, 
\bea                                  \label{GL}
\stackrel{(0)}{\Box} h_{\mu\nu}+2\stackrel{(0)}{R}_{\mu\alpha\nu\beta}h^{\alpha\beta}=
h_{\mu\nu},~~~~~~
\stackrel{(0)}{\nabla}_\mu h^\mu_\nu=0,~~~h^\mu_\mu=0. 
\eea
Setting 
$
h_{\mu\nu}=e^{i{\omega} t}H_{\mu\nu}(r,\vartheta,\varphi)
$
they admit a bound state solution with 
${\omega^2}<0$,
provided that the black hole radius 
$r_h<0.86$ \cite{Brito:2013wya}. 
It follows that small black holes are unstable
since $\omega$ is imaginary and the perturbations grow in time 
\cite{Babichev:2013una}. The condition on $r_h$ is not crucial since 
the lengthscale is of the order 
of the Hubble radius, hence all ordinary  black holes are small enough 
to be unstable. On the other hand, the instability is very mild and takes a Hubble time  to develop. 
Therefore, if real astrophysical black holes were described by the bigravity
theory, their instability would be largely irrelevant and they would actually 
be stable for all practical purposes over a cosmologically long period of time.  

Additional information on black holes in theories with massive gravitons can be found in the 
review articles \cite{Volkov:2014ooa,Babichev:2015xha}.

\section{Concluding remarks}

Above is only a very brief description of stationary and asymptotically flat hairy black holes in 3+1 spacetime 
dimensions. One finds many more of them  if one abandons the asymptotic flatness condition, 
if one considers different horizon topologies, if one goes to higher spacetime dimensions, etc. 
In view of this variety of hairy black holes one can wonder if there is still any sense in the no-hair conjecture~? 

Surprizingly, the answer seems to be affirmative  for the astrophysically relevant, which means stable and 
macroscopically large  black holes. This simply follows from the fact that Einstein-Maxwell
theory is definitely valid at the macroscopic scale, but within this theory the no-hair conjecture is proven. 
The situation is different at the microscopic scale where other theories apply, as for example
the non-Abelian models. There are stable hairy black holes in that case, for example those supporting the
Yang-Mills-Higgs or Skyrme hair. However, they are always microscopically small
and loose their hair when grow beyond a certain size.  

If there exists  a  cosmic scalar field accounting for the dark energy, 
this might perhaps modify the macroscopic black hole structure, however, most of the 
known solutions with scalar hair are unstable. 
Black holes in models violating the energy conditions are unstable, 
while the known non-minimally coupled Horndeski solutions \eqref{SS} and \eqref{rinal} are either 
unstable or non-asymptotically flat. However, the Horndeski theory is quite general and  it is 
not excluded that it may admit stable hairy black holes.  The possible candidates are 
black holes \cite{Mignemi:1992nt,Kanti:1995vq,Torii:1996yi,Alexeev:1996vs} in the dilaton-Gauss-Bonnet theory \eqref{EYMD} 
 (and maybe their 
cousins \cite{Sotiriou:2014pfa} in the 
Galileon theory \eqref{GB}) because they do not have the most dangerous  
S and P sector instabilities \cite{Kanti:1995vq,Pani:2009wy}. 
Notice however that their hair is only secondary. 

If theories with massive gravitons indeed describe the world, then the black holes are almost 
Kerr-Newman, apart from a narrow region in the horizon vicinity where a slow 
accretion of massive graviton modes takes place. 

In summary, it seems that all  stationary black holes that are astrophysically relevant 
indeed obey the no-hair conjecture, 
at least no explicit counter-examples are known.


\end{document}